\documentclass[lettersize,journal]{IEEEtran}
\usepackage{amsmath,amsfonts}
\usepackage{algorithmic}
\usepackage{algorithm}
\usepackage{array}
\usepackage[caption=false,font=normalsize,labelfont=sf,textfont=sf]{subfig}
\usepackage{textcomp}
\usepackage{stfloats}
\usepackage{url}
\usepackage{verbatim}
\usepackage{graphicx}
\usepackage{cite}
\usepackage{pifont}%
\usepackage{xcolor}
\usepackage{bm}
\makeatletter
\def\ps@IEEEtitlepagestyle{%
  \def\@oddfoot{\mycopyrightnotice}%
  \def\@evenfoot{}}
\def\mycopyrightnotice{%
  \hfill \footnotesize
  Accepted in: \textit{IEEE Communications Letters}, Early Access, Sept. 22, 2025. 
  DOI: \texttt{10.1109/LCOMM.2025.3612448}\hfill
}
\makeatother

\begin{document}

\title{Robust Energy-Efficient DRL-Based Optimization in UAV-Mounted RIS Systems with Jitter}

\author{Mahmoud M. Salim, Khaled M. Rabie, \IEEEmembership{(Senior Member, IEEE)}, and Ali H. Muqaibel,
\IEEEmembership{(Senior Member, IEEE)}        
\thanks{All authors are with the Center for Communication Systems and Sensing, King Fahd University of Petroleum and Minerals, Dhahran 31261, Saudi Arabia. Also, Khaled M. Rabie is affiliated with the Computer Engineering Department, and Ali H. Muqaibel is with the Electrical Engineering Department at King Fahd University of Petroleum and Minerals, Dhahran 31261, Saudi Arabia. Mahmoud M. Salim is the corresponding author.}}

\markboth{Journal of \LaTeX\ Class Files,~Vol.~14, No.~8, August~2021}%
{Shell \MakeLowercase{\textit{et al.}}: A Sample Article Using IEEEtran.cls for IEEE Journals}


\maketitle

\begin{abstract}
In this letter, we propose an energy-efficient design for an unmanned aerial vehicle (UAV)-mounted reconfigurable intelligent surface (RIS) communication system with nonlinear energy harvesting (EH) and UAV jitter. A joint optimization problem is formulated to maximize the EH efficiency of the UAV-mounted RIS by controlling the user powers, RIS phase shifts, and time-switching factor, subject to quality of service and practical EH constraints. The problem is nonconvex and time-coupled due to UAV angular jitter and nonlinear EH dynamics, making it intractable for conventional optimization methods. To address this, we reformulate the problem as a deep reinforcement learning (DRL) environment and develop a smoothed softmax dual deep deterministic policy gradient algorithm. The proposed method incorporates action clipping, entropy regularization, and softmax-weighted Q-value estimation to improve learning stability and exploration. Simulation results show that the proposed algorithm converges reliably under various UAV jitter levels and achieves an average EH efficiency of 45.07\%, approaching the 53.09\% upper bound of exhaustive search, and outperforming other DRL baselines.
\end{abstract}

\begin{IEEEkeywords}
UAV, RIS, DRL, jitter, energy harvesting.
\end{IEEEkeywords}

\section{Introduction and Related Work} \label{Sec_Intro}
Unmanned aerial vehicles (UAVs) have emerged as agile and cost-effective platforms for enhancing wireless connectivity in challenging or temporary deployment scenarios \cite{wang-2023,pandey2024uav}. When integrated with reconfigurable intelligent surfaces (RISs), UAVs can intelligently control signal propagation to enable energy-efficient and line-of-sight (LoS) dominant communication links \cite{ahmed2025comprehensive}. UAV-mounted RIS systems support adaptive beamforming and extended coverage, particularly in dense urban and emergency environments \cite{zhang-2023-joint,xu-2024-ris}.

UAV-mounted RISs have been studied in various architectures, including amplify-and-forward relaying \cite{zhang-2023-joint} and industrial IoT support \cite{xu-2024-ris}, using both fixed-wing and rotary-wing platforms to improve coverage and endurance \cite{lin-2024-green}. However, a persistent limitation is the short operational time of UAV-mounted RIS systems, driven by onboard energy constraints. To address this, RF energy harvesting (EH) has been incorporated into UAV-RIS system designs \cite{salim2025energy}. Robust deep reinforcement learning (DRL)-based EH optimization was proposed in \cite{peng-2023}, and a twin-delayed deep deterministic policy gradient (TD3)-based framework was introduced in \cite{puspitasari-2024-td3} for joint EH and information transmission.

A critical challenge in UAV-mounted RIS systems is UAV jitter, referring to random angular deviations caused by wind or mechanical instabilities, which misaligns the RIS and degrades beamforming and EH performance \cite{wang-2024-aerial, yu-2023-aerial}. While recent studies have addressed RIS configuration and UAV positioning under jitter \cite{liu-2025-throughput}, the presence of nonlinear EH behavior further complicates the problem. This leads to a complex, time-coupled optimization problem of different parameters such as transmit power, RIS phase shifts, and EH scheduling. Conventional optimization methods, such as convex relaxation or alternating optimization, are inadequate to handle these dynamics in real environments \cite{li-2023-joint}. Although recent DRL-based approaches have been applied to similar scenarios \cite{peng-2023, puspitasari-2024-td3}, they often neglect the joint modeling of UAV jitter and nonlinear EH, which limits their reliability and adaptability in realistic UAV-RIS system deployments.

In contrast, this letter proposes a DRL-based framework for the joint optimization of user powers, RIS phase shifts, and time-switching (TS) ratio, explicitly accounting for angular jitter, nonlinear RF EH behavior, and user quality of service (QoS) requirements. To the best of our knowledge, this is the first DRL-based solution designed to maximize nonlinear EH efficiency in UAV-mounted RIS systems under UAV jitter. The contributions of this letter are summarized as follows: 1) the nonconvex, time-coupled optimization problem is reformulated as a Markov decision process (MDP) for the control of user powers, RIS configuration, and TS allocation; 2) a smoothed softmax dual deep deterministic policy gradient (SSD3) algorithm is developed to solve this MDP-based optimization problem, integrating dual actor-critic networks, clipped noise for exploration, softmax-weighted Q-value estimation, and entropy-regularized actor updates; 3) simulation results show that SSD3 outperforms deep deterministic policy gradient (DDPG) and TD3 baselines, achieving higher EH efficiency and faster convergence under practical UAV jitter.

\section{System Model}\label{Sec_SM}
\begin{figure}[!t]
\centering
\includegraphics[width=0.8\columnwidth]{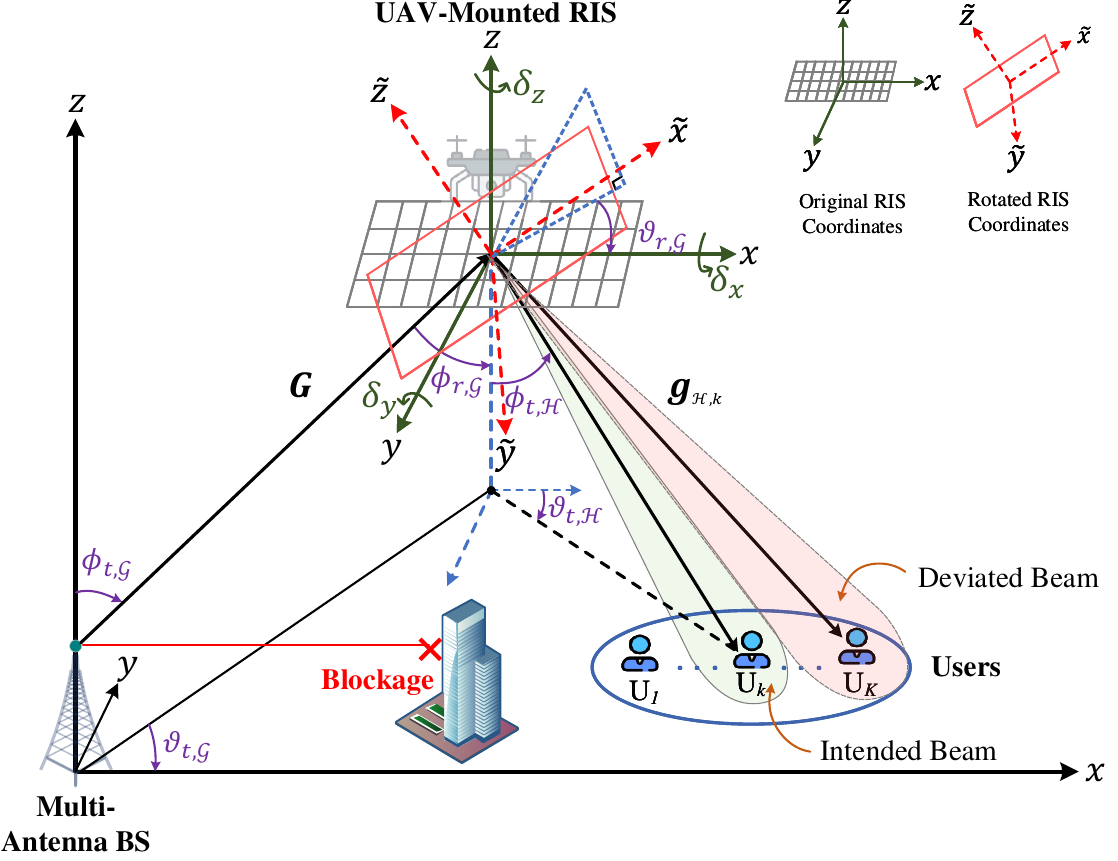}
\caption{Proposed UAV-mounted RIS system model with jitter.}
\label{Fig_SM}
\end{figure}
We consider the UAV-mounted RIS wireless communication system illustrated in Fig.~\ref{Fig_SM}, where a ground base station (BS) located at \( \mathbf{q}_{\mathrm{BS}} = [0, 0, H_{\mathrm{BS}}]^T \) is equipped with \( M = M_y M_z \) antennas, spaced by \( \Delta_y^{\mathrm{BS}} \) and \( \Delta_z^{\mathrm{BS}} \) along the horizontal and vertical axes, respectively. The BS serves \( K \) single-antenna ground users with locations given by \( \mathbf{q}_{\mathrm{U}_k} = [x_k, y_k, 0]^T \), \( k = 1, \dots, K \) assisted by a UAV-mounted RIS, which is implemented as a uniform planar array with \( N = N_x N_y \) passive reflecting elements. The UAV hovers at a fixed altitude \( H_{\mathrm{RIS}} \), with center coordinate \( \mathbf{q}_{\mathrm{RIS}} = [x_{\mathrm{RIS}}, y_{\mathrm{RIS}}, H_{\mathrm{RIS}}]^T \). Each RIS element is uniformly spaced with inter-element distances \( \Delta_x^{\mathrm{RIS}} \) and \( \Delta_y^{\mathrm{RIS}} \) along the planar array. We assume far-field propagation for both BS–RIS and RIS–$U_k$ links, enabling angular modeling via azimuth and elevation angles \cite{yu-2023-aerial}. The angle of departure (AoD) from the BS is denoted as \( (\phi_{t,\mathcal{G}}, \vartheta_{t,\mathcal{G}}) \), and the angle of arrival (AoA) at the RIS is \( (\phi_{r,\mathcal{G}}, \vartheta_{r,\mathcal{G}}) \). The AoD from the RIS to the \( k \)-th user is given by \( (\phi_{t,\mathcal{H}}, \vartheta_{t,\mathcal{H}}) \). The transmit array response at the BS is modeled as the Kronecker product of its vertical and horizontal steering vectors with respect to the AoD as

\vspace{-0.05in}
{\footnotesize
\begin{align}
\boldsymbol{\alpha}_{t,\mathcal{G}} &= 
\left[1, \ldots, e^{-j \frac{2\pi \Delta_y^{\mathrm{BS}}}{\lambda} (M_y - 1) \sin(\phi_{t,\mathcal{G}})\cos(\vartheta_{t,\mathcal{G}})} \right]^T \nonumber \\
&\otimes \left[1, \ldots, e^{-j \frac{2\pi \Delta_z^{\mathrm{BS}}}{\lambda} (M_z - 1) \sin(\phi_{t,\mathcal{G}})\sin(\vartheta_{t,\mathcal{G}})} \right]^T,
\end{align}}where $\lambda$ denotes the carrier wavelength. Similarly, the receiving array response at the UAV-mounted RIS expressed using the AoA is

\vspace{-0.05in}
{\footnotesize
\begin{align}
\boldsymbol{\alpha}_{r,\mathcal{G}} &= 
\left[1, \ldots, e^{-j \frac{2\pi \Delta_x^{\mathrm{RIS}}}{\lambda} (N_x - 1) \sin(\phi_{r,\mathcal{G}})\cos(\vartheta_{r,\mathcal{G}})} \right]^T \nonumber \\
&\otimes \left[1, \ldots, e^{-j \frac{2\pi \Delta_y^{\mathrm{RIS}}}{\lambda} (N_y - 1) \sin(\phi_{r,\mathcal{G}})\sin(\vartheta_{r,\mathcal{G}})} \right]^T.
\end{align}}

Similarly, the transmit array response toward user $k$, defined by its AoD, can be calculated and denoted as $\boldsymbol{\alpha}_{t,\mathcal{H}}$.

\subsection{Channel Model}
We adopt an air-to-ground (ATG) channel model for the BS–RIS link \cite{alhourani-2014-optimal,yu-2023-aerial}, and a Rician fading model for the RIS–$U_k$ link \cite{peng-2023}. The BS-to-RIS channel matrix \( \mathbf{G} \in \mathbb{C}^{N \times M} \) comprises both LoS and non-line-of-sight (NLoS) components, modeled as

\vspace{-0.05in}
{\footnotesize
\begin{align}
\mathbf{G} = \sqrt{\beta_\mathcal{G}} \left( 
\sqrt{P_{\mathrm{LoS}}} \mathbf{H}_{\mathcal{G}}^{\mathrm{LoS}}  
+ \sqrt{1 - P_{\mathrm{LoS}}} \mathbf{H}_{\mathcal{G}}^{\mathrm{NLoS}} \right),
\label{eq_G}
\end{align}}where \( \beta_\mathcal{G} = \beta_0 / d_\mathcal{G}^2 \) denotes large-scale path loss, with \( \beta_0 \) as the reference path loss and \( d_\mathcal{G} = \| \mathbf{q}_{\mathrm{RIS}} - \mathbf{q}_{\mathrm{BS}} \| \). The deterministic LoS component is given by \( \mathbf{H}_{\mathcal{G}}^{\mathrm{LoS}} = e^{-j \frac{2\pi d_\mathcal{G}}{\lambda}} \boldsymbol{\alpha}_{r,\mathcal{G}} \boldsymbol{\alpha}_{t,\mathcal{G}}^H \), where the exponential term accounts for the propagation delay. The NLoS term \( \mathbf{H}_{\mathcal{G}}^{\mathrm{NLoS}} \in \mathbb{C}^{N \times M} \) follows i.i.d. Rayleigh fading with entries \( \mathcal{CN}(0,1) \) \cite{al2014modeling}.

Based on the ATG model in \cite{alhourani-2014-optimal}, the LoS probability is given by

\vspace{-0.05in}
{\footnotesize
\begin{equation}
P_{\mathrm{LoS}} = \frac{1}{1 + a \exp\left(-b(\phi_{t,\mathcal{G}} - a)\right)},
\end{equation}}where the elevation angle $\phi_{t,\mathcal{G}} = \frac{180}{\pi} \sin^{-1} \left( \frac{H_{\mathrm{RIS}} - H_{\mathrm{BS}}}{d_\mathcal{G}} \right)$, while \( a \) and \( b \) are environment-dependent empirical constants \cite{wang-2024-age}.

For the RIS-to-$U_k$ link, the large-scale path loss is \( \beta_{\mathcal{H},k} = \beta_0 / d_{\mathcal{H},k} \), with \( d_{\mathcal{H},k} = \| \mathbf{q}_{\mathrm{RIS}} - \mathbf{q}_{\mathrm{U}_k} \| \). Thus, the small-scale fading is modeled as

\vspace{-0.05in}
{\footnotesize
\begin{align}
\mathbf{g}_{\mathcal{H},k} = \sqrt{\beta_{\mathcal{H},k}} \left( 
\sqrt{\frac{K_\mathcal{H}}{1 + K_\mathcal{H}}} \mathbf{h}_{\mathcal{H},k}^{\mathrm{LoS}} + 
\sqrt{\frac{1}{1 + K_\mathcal{H}}} \mathbf{h}_{\mathcal{H},k}^{\mathrm{NLoS}} 
\right),
\label{eq_g}
\end{align}
}where \( K_\mathcal{H} \) is the Rician factor. The LoS component is \( \mathbf{h}_{\mathcal{H},k}^{\mathrm{LoS}} = e^{-j \frac{2\pi d_{\mathcal{H},k}}{\lambda}} \boldsymbol{\alpha}_{t,\mathcal{H}} \in \mathbb{C}^{N \times 1} \), and the NLoS component \( \mathbf{h}_{\mathcal{H},k}^{\mathrm{NLoS}} \in \mathbb{C}^{N \times 1} \) is Rayleigh fading with i.i.d. \( \mathcal{CN}(0,1) \) entries.
\subsection{Time-Switching Energy Harvesting Model}
We adopt a TS protocol to manage the trade-off between EH and information reflection in the UAV-mounted RIS system. Each time slot \( t \) is divided into two non-overlapping phases: an EH phase of duration \( \tau(t) \) and a transmission phase of duration \( 1 - \tau(t) \), where \( \tau(t) \in [0,1] \) is the TS factor. The BS transmits a composite signal \( \mathbf{S}(t) = \sum_{k \in \mathcal{K}} \mathbf{V}_k \mathcal{C}_k \), where \( \mathbf{V}_k \in \mathbb{C}^{M \times 1} \) is the beamforming vector for user \( k \), and \( \mathcal{C}_k \sim \mathcal{CN}(0,1) \) is the data symbol. The total BS transmit power satisfies \( \mathbb{E}[\mathbf{S}^H \mathbf{S}] = \sum_{k \in \mathcal{K}} \|\mathbf{V}_k\|^2 \).

During the EH phase, all RIS elements harvest power from the BS-transmitted signal. The ideal received RF power at the RIS is given by

\vspace{-0.05in}
{\footnotesize
\begin{equation}
P^{\text{RF}}(t) = \tau(t) \sum_{i=1}^{N_x} \sum_{j=1}^{N_y} \left\| \mathbf{g}_{i,j}^H \mathbf{S}(t) \right\|^2,
\end{equation}
}where \( \mathbf{g}_{i,j} \in \mathbb{C}^{M \times 1} \) is the channel vector from the BS to RIS element \( (i, j) \), extracted from the BS-to-RIS channel matrix \( \mathbf{G} \in \mathbb{C}^{N \times M} \). To account for circuit nonlinearity, the actual harvested power follows a nonlinear model \cite{sharma-2022,Semiha2023}

{\footnotesize
\begin{equation}
\varepsilon^{\text{RF-NL}}(t) = \frac{\Omega(t) - P_{\text{sat}} \Delta}{1 - \Delta},
\label{eq:EH_model}
\end{equation}
}where {\footnotesize \( \Omega(t) = P_{\text{sat}} / (1 + \exp(-c(P^{\text{RF}}(t) - d))) \)} and {\footnotesize \( \Delta = 1 / (1 + \exp(cd)) \)}. Here, \( P_{\text{sat}} \) denotes the saturation power, and \( c \), \( d \) are nonlinearity parameters subject to harvester hardware design \cite{salim-2024}.

\subsection{Transmission and Received Signal Model}
In the transmission phase, RIS elements reflect the incoming signal, where the received signal at user \( k \) is

\vspace{-0.05in}
{\footnotesize
\begin{equation}
y_k = \mathbf{g}_{\mathcal{H},k}^H \boldsymbol{\Theta} \mathbf{G} \mathbf{S}(t) + n_k,
\end{equation}
}where \( \boldsymbol{\Theta} = \operatorname{diag}(e^{j\theta_1}, \dots, e^{j\theta_N}) \) is the RIS phase-shift matrix, and \( n_k \sim \mathcal{CN}(0, \sigma_k^2) \) is the additive white Gaussian noise. Assuming ideal interference cancellation \cite{peng-2023}, the SNR at user \( k \) is

\vspace{-0.05in}
{\footnotesize
\begin{equation}
\Gamma_k(t) = \frac{\left| \mathbf{g}_{\mathcal{H},k}^H \boldsymbol{\Theta} \mathbf{G} \mathbf{V}_k \right|^2}{\sum\limits_{\substack{u \in \mathcal{K} \\ u \neq k}} \left| \mathbf{g}_{\mathcal{H},k}^H \boldsymbol{\Theta} \mathbf{G} \mathbf{V}_u \right|^2 + \sigma_k^2},
\end{equation}
}

Therefore, the achievable rate (bps/Hz) for user \( k \) in slot \( t \) is

\vspace{-0.05in}
{\footnotesize
\begin{equation}
R_k(t) = (1 - \tau(t)) B \log_2 \left(1 + \Gamma_k(t) \right), \quad k \in \mathcal{K},\ t \in \mathcal{T}.
\end{equation}
}

Hence, the overall EH efficiency at the UAV-mounted RIS is defined as

\vspace{-0.05in}
{\footnotesize
\begin{equation}
\varepsilon\varepsilon(t) = \frac{\varepsilon^{\text{RF-NL}}(t)}{\varepsilon^{\text{r}}_{\text{UAV-RIS}}(t)},
\end{equation}
}where {\footnotesize \( \varepsilon^{\text{r}}_{\text{UAV-RIS}}(t) = \sum\limits_{i=1}^{N_x} \sum\limits_{j=1}^{N_y} \left\| \mathbf{g}_{i,j}^H \mathbf{S}(t) \right\|^2 \)} denotes the total received RF power at the RIS. This formulation captures the conversion efficiency under practical nonlinear EH dynamics.

\subsection{UAV-mounted RIS Jitter Modeling}
In practical deployments, the UAV-mounted RIS may undergo random angular jitter due to wind, vibrations, or mechanical instabilities, causing orientation deviations \cite{wang-2024-aerial}. These misalignments alter the signal incidence angles at the RIS, degrading EH efficiency and beamforming performance. Hence, accurately modeling jitter is essential to optimize the EH–reflection trade-off. We assume the RIS lies on the horizontal \( xy \)-plane. Under jitter, the RIS plane rotates to a new orientation \( \tilde{x} \tilde{y} \). While the BS remains static, the AoAs and AoDs at the RIS vary with the UAV’s angular displacement. Jitter is modeled as small-angle deviations \( \delta_x, \delta_y, \delta_z \sim \mathcal{N}(0, \sigma_j^2) \) around the roll, pitch, and yaw axes \cite{yu-2023-aerial}. These induce a composite 3D rotation matrix {\footnotesize $\mathbf{R}_j = \mathbf{R}_{\text{yaw}}(\delta_z)\, \mathbf{R}_{\text{pitch}}(\delta_y)\, \mathbf{R}_{\text{roll}}(\delta_x)$}, where

\vspace{-0.05in}
{\footnotesize
\begin{align}
\mathbf{R}_{\text{yaw}}(\delta_z) &= 
\begin{bmatrix}
\cos\delta_z & -\sin\delta_z & 0 \\
\sin\delta_z & \cos\delta_z  & 0 \\
0            & 0             & 1
\end{bmatrix}, \\
\mathbf{R}_{\text{pitch}}(\delta_y) &= 
\begin{bmatrix}
\cos\delta_y  & 0 & \sin\delta_y \\
0             & 1 & 0 \\
-\sin\delta_y & 0 & \cos\delta_y
\end{bmatrix}, \\
\mathbf{R}_{\text{roll}}(\delta_x) &= 
\begin{bmatrix}
1 & 0           & 0 \\
0 & \cos\delta_x & -\sin\delta_x \\
0 & \sin\delta_x & \cos\delta_x
\end{bmatrix}.
\end{align}
}

Accordingly, the effective geometric vectors from the BS to the RIS and from the RIS to user \( k \), under jitter, become

\vspace{-0.05in}
{\footnotesize
\begin{align}
\widetilde{\mathbf{v}}_{\mathcal{G}} &= \mathbf{R}_j (\mathbf{q}_{\mathrm{RIS}} - \mathbf{q}_{\mathrm{BS}}), \\
\widetilde{\mathbf{v}}_{\mathcal{H},k} &= \mathbf{R}_j (\mathbf{q}_{\mathrm{U},k} - \mathbf{q}_{\mathrm{RIS}}).
\end{align}
}

The resulting AoAs and AoDs at RIS are recalculated as

\vspace{-0.05in}
{\footnotesize
\begin{align}
\widetilde{\vartheta}_{r,\mathcal{G}} &= \arccos \left( \frac{ \langle \widetilde{\mathbf{v}}_{\mathcal{G}}, \mathbf{e}_z \rangle }{ \| \widetilde{\mathbf{v}}_{\mathcal{G}} \| \, \| \mathbf{e}_z \| } \right), \\
\widetilde{\varphi}_{r,\mathcal{G}} &= \arccos \left( \frac{ \langle \mathcal{P}_{xy}(\widetilde{\mathbf{v}}_{\mathcal{G}}), \mathbf{e}_x \rangle }{ \| \mathcal{P}_{xy}(\widetilde{\mathbf{v}}_{\mathcal{G}}) \| \, \| \mathbf{e}_x \| } \right), \\
\widetilde{\vartheta}_{t,\mathcal{H}} &= \arccos \left( \frac{ \langle \widetilde{\mathbf{v}}_{\mathcal{H},k}, -\mathbf{e}_z \rangle }{ \| \widetilde{\mathbf{v}}_{\mathcal{H},k} \| \, \| \mathbf{e}_z \| } \right), \\
\widetilde{\varphi}_{t,\mathcal{H}} &= \arccos \left( \frac{ \langle \mathcal{P}_{xy}(\widetilde{\mathbf{v}}_{\mathcal{H},k}), \mathbf{e}_x \rangle }{ \| \mathcal{P}_{xy}(\widetilde{\mathbf{v}}_{\mathcal{H},k}) \| \, \| \mathbf{e}_x \| } \right),
\end{align}
}where \( \mathbf{e}_x \) and \( \mathbf{e}_z \) are unit vectors along the \( x \)- and \( z \)-axes, \( \mathcal{P}_{xy}(\cdot) \) denotes the orthogonal projection onto the \( xy \)-plane, and \( \langle \cdot, \cdot \rangle \) is the Euclidean inner product. These perturbed angles are then used to update the channel components in \eqref{eq_G} and \eqref{eq_g}, thereby incorporating the impact of UAV jitter.

\section{Problem Formulation} \label{Sec_PF}

Based on the system model in Section~\ref{Sec_SM}, we aim to maximize the EH efficiency of the UAV-mounted RIS system over a finite horizon of \( T \) time slots as \( \mathbb{T} = \{1, 2, \dots, T\} \), while ensuring minimum QoS for all users. This is achieved by jointly optimizing the BS power allocation, the RIS phase shifts, and the TS factor at each slot \( t \). The resulting optimization problem is formulated as

\vspace{-0.05in}
{\footnotesize
\begin{IEEEeqnarray}{cl}
\IEEEyesnumber\label{eq:MOF_TS} \IEEEyessubnumber*
\max_{\tau(t), \mathbf{P}_U, \bm{\Theta}} & \sum_{t=1}^{T} \varepsilon\varepsilon(t) \\
\text{s.t.} \quad & R_k(t) \geq R_{\min},\quad \forall k \in \mathbb{K},\ t \in \mathbb{T}, \label{eq:MOF_TS_rate}\\
& 0 \leq \tau(t) \leq 1,\quad \forall t \in \mathbb{T}, \\
& 0 \leq P_{BS}(t) \leq P_{BS}^{\max},\quad \forall t \in \mathbb{T}, \\
& 0 \leq p_k(t) \leq P_U^{\max},\quad \forall k \in \mathbb{K},\ t \in \mathbb{T}, \\
& \theta_n(t) \in [0, 2\pi],\quad \forall n \in [1, N],\ t \in \mathbb{T}, \\
& |e^{j\theta_n(t)}| = 1,\quad \forall n \in [1, N],\ t \in \mathbb{T}\label{eq:MOF_TS_Unit},
\end{IEEEeqnarray}
}where $\tau(t)$ is the TS factor, \( \mathbf{P}_U = [p_1(t), \dots, p_K(t)] \) and \( \bm{\Theta} = [\theta_1(t), \dots, \theta_N(t)] \) represent the user-specific BS power allocations and RIS phase shifts, respectively. Constraints \eqref{eq:MOF_TS_rate}–\eqref{eq:MOF_TS_Unit} guarantee user QoS, feasible TS allocation, BS and user power limits, valid phase shift bounds for RIS elements, and the unit-modulus property required for passive reflection, respectively.

\section{Proposed SSD3-Based Optimization Framework} \label{Sec_Proposed_Algorithm}

The optimization problem in \eqref{eq:MOF_TS} is non-convex and time-coupled, involving continuous control variables under nonlinear constraints. In particular, the TS ratio, RIS phase shifts, and user-specific BS power allocations are interdependent and evolve over channels affected by UAV jitter and nonlinear EH dynamics. Conventional methods, such as convex relaxation or alternating optimization, are not well suited for real-time adaptation. To address this, we develop a DRL framework based on the DDPG algorithm.
\subsection{MDP Formulation}
We model the system as a MDP defined by the tuple \( \langle \mathcal{S}, \mathcal{A}, P_t, R, \gamma \rangle \), where \( \mathcal{S} \) and \( \mathcal{A} \) denote the state and action spaces, \( P_t \) is the transition probability, \( R \) is the reward function, and \( \gamma \in [0,1) \) is the discount factor. The state at time slot \( t \) is

\vspace{-0.05in}
{\footnotesize
\begin{equation}
s^t = \big[ \Re\{\mathbf{G}\}, \Im\{\mathbf{G}\}, \Re\{\mathbf{g}_{\mathcal{H},k}\}, \Im\{\mathbf{g}_{\mathcal{H},k}\}, \mathcal{P}^{\mathrm{RIS}}_{i,j} , \mathcal{P}_k, a^{t-1} \big],
\end{equation}}where \( \mathcal{P}^{\mathrm{RIS}}_{i,j} \) denotes the coordinates of the \( (i,j) \)-th RIS element, \( \mathcal{P}_k \) is the user location, and \( a^{t-1} \) is the previous action. The action vector is \( a^t = [\tau(t), p_k(t), \theta_n(t)] \), consisting of the TS ratio, BS power allocation, and RIS phase shifts. The reward \( r^t \) is set to the EH efficiency \( \varepsilon\varepsilon(t) \) if users' QoS constraints are satisfied; otherwise, it is zero.

In DDPG, an actor network maps states to actions, and a critic evaluates Q-values. The actor is trained using the deterministic policy gradient

\vspace{-0.05in}
{\footnotesize
\begin{equation}
    \nabla_{\vartheta^{\mu}} J \approx \frac{1}{N_b} \sum_i \nabla_a Q(s, a | \vartheta^Q) |_{a=\mu(s_i)} \nabla_{\vartheta^{\mu}} \mu(s_i),
\end{equation}
}where \( N_b \) is the mini-batch size, and $\vartheta^{\mu}$ and $\vartheta^{Q}$ are the parameters of the actor and critic networks, respectively. The critic minimizes the loss

\vspace{-0.05in}
{\footnotesize
\begin{equation}
L(\vartheta^Q) = \frac{1}{N_b} \sum_i (y_i - Q(s_i, a_i | \vartheta^Q))^2,
\end{equation}
}with target value \( y_i = r_i + \gamma Q'(s_{i+1}, \mu'(s_{i+1})) \). Here, \( Q' \) and \( \mu' \) denote the target critic and actor networks, respectively. Soft target updates are applied as 

\vspace{-0.05in}
{\footnotesize
\begin{equation}
\vartheta^{\mu'} \leftarrow \rho \vartheta^{\mu} + (1 - \rho) \vartheta^{\mu'}, \quad
\vartheta^{Q'} \leftarrow \rho \vartheta^Q + (1 - \rho) \vartheta^{Q'}.
\end{equation}
}

Exploration during training is encouraged by adding Gaussian noise to the actor’s output as \( a = \mu(s) + \xi \), with \( \xi \sim \mathcal{N}(0, \sigma^2) \).
\subsection{Smoothed Softmax Dual DDPG (SSD3) Algorithm}
While DDPG is effective for continuous control, it tends to overestimate Q-values due to a single critic. TD3 mitigates this by using two critics and computing the target value as

\vspace{-0.05in}
{\footnotesize
\begin{equation}
    y_i = r_i + \gamma \min\{ Q'_1(s_{i+1}, \mu'(s_{i+1})), Q'_2(s_{i+1}, \mu'(s_{i+1})) \},
\end{equation}
}with delayed actor updates. However, this conservative approach can lead to underestimation and slower convergence.

To address this, we propose the SSD3 algorithm detailed in Algorithm \ref{alg:SSD3}, which extends TD3 by employing two independent actors and using a softmax-weighted critic fusion instead of the conservative minimum. SSD3 further introduces entropy regularization in the actor update to enhance exploration and stability under the nonconvex and stochastic dynamics of UAV-RIS systems. Specifically, SSD3 maintains two actor–critic pairs \((\mu_1, Q_1)\) and \((\mu_2, Q_2)\). For target estimation, clipped Gaussian noise \( \epsilon \sim \mathcal{N}(0, \sigma^2) \) is added to the target action, forming \( a'^i = \mu'(s') + \text{clip}(\epsilon, -c, c) \). The softmax-weighted target is computed as:

\vspace{-0.05in}
{\footnotesize
\begin{equation}
T_{\mathrm{SSD3}}(s') = \frac{\sum_i \exp(\beta Q_{\min}^i) Q_{\min}^i}{\sum_i \exp(\beta Q_{\min}^i)},
\end{equation}
}where \( Q_{\min}^i = \min\{ Q_1'(s', a'^i), Q_2'(s', a'^i) \} \), and \( \beta \) controls the softmax temperature. To reduce bias, SSD3 dynamically selects the actor corresponding to the higher Q-value at each training step

\vspace{-0.05in}
{\footnotesize
\begin{equation}
\mu_{\text{chosen}} = 
\begin{cases}
\mu_1, & \text{if } Q_1(s, \mu_1(s)) \geq Q_2(s, \mu_2(s)), \\
\mu_2, & \text{otherwise},
\end{cases}
\end{equation}
}
and updates it using an entropy-regularized loss function

\vspace{-0.05in}
{\footnotesize
\begin{equation}
\mathcal{L}_{\mu_{\text{chosen}}} = -\mathbb{E}_{s \sim \mathcal{D}} \left[ Q_{\max}(s, \mu_{\text{chosen}}(s)) \right] 
+ \lambda_{\mathrm{ent}} \cdot \mathbb{E}_{s} \left[ \|\mu_{\text{chosen}}(s)\|_2^2 \right],
\end{equation}
}where \( Q_{\max}(s, a) = \max\{ Q_1(s, a), Q_2(s, a) \} \), and \( \lambda_{\mathrm{ent}} \) controls the strength of entropy-based regularization.
\begin{algorithm}[!t]
\caption{SSD3-Based UAV-RIS Optimization Algorithm}
\label{alg:SSD3}
\begin{algorithmic}[1]
\scriptsize
\STATE \textbf{Initialize:} dual actor-critic networks and their target networks
\FOR{each episode}
    \STATE Reset environment and observe initial state $s^0$
    \FOR{each step $t$}
        \STATE Select actor based on critic scores; take action $a^t$
        \STATE Apply control: TS ratio, BS power, RIS phase shifts
        \STATE Observe reward $r^t$, next state $s^{t+1}$; store transition
        \IF{policy update time}
            \STATE Update critics using softmax-weighted Q-target
            \STATE Update selected actor with entropy-regularized loss
            \STATE Soft update target networks
        \ENDIF
    \ENDFOR
\ENDFOR
\STATE \textbf{Output:} Optimal policy $\mu^*(s)$ for $\{\tau^*, \mathbf{P}_U^*, \bm{\Theta}^*\}$
\end{algorithmic}
\end{algorithm}

In terms of computational complexity, we focus on the online deployment phase, assuming that training is performed offline. During deployment, the proposed SSD3 agent selects actions through a forward pass of the chosen actor network. For an actor network with \( L \) hidden layers, each containing \( n \) neurons, and input and output dimensions \( I \) and \( O \), respectively, the worst-case per-step complexity is \( \mathcal{O}(\max(In, n^2, nO)) \). The use of entropy-regularized actor updates and softmax-weighted Q-targets improves training robustness but does not affect inference-time complexity. Compared to iterative optimization methods, SSD3 enables real-time decision-making with significantly lower computational overhead at deployment.

\section{Simulation Results} \label{Sec_Simu}
This section evaluates the performance of the proposed SSD3 algorithm for a UAV-mounted RIS system with \( K = 3 \) users. We examine its convergence under varying UAV jitter levels and compare its EH efficiency against TD3, DDPG, and exhaustive search. The results demonstrate the robustness and near-optimality of SSD3. Table~\ref{tab:Simu} lists the simulation parameters used unless stated otherwise.

\begin{table}[!t]
\caption{Simulation Parameters}
\label{tab:Simu}
\centering
\scalebox{0.8}{
\begin{tabular}{| c | c | c |}
\hline
\textbf{Symbol} & \textbf{Description} & \textbf{Value} \\
\hline
$A_{\text{sim}}$ & Simulation area & $20$ m $\times$ $20$ m \\
\hline
$N$ & Number of RIS elements & $16$ \\
\hline
$a$, $b$ & LoS environment constants & $9.61$, $0.16$ \\
\hline
$\beta_0$ & Reference path loss at $d_{\text{ref}} = 1$ m & $-30$ dB \\
\hline
$\sigma_k^2$ & Noise power at user $k$ & $-102$ dBm \\
\hline
$P_{BS}^{\max}$ & BS maximum transmit power & $500$ W \\
\hline
$R_{\text{min}}$ & Minimum user rate requirement & $70$ Mbps \\
\hline
$c$, $d$ & Non-linear harvested parameters& $6400$, $0.003$ \\
\hline
\end{tabular}}
\end{table}

\begin{figure}[!t]
\centering
\includegraphics[width=0.7\columnwidth]{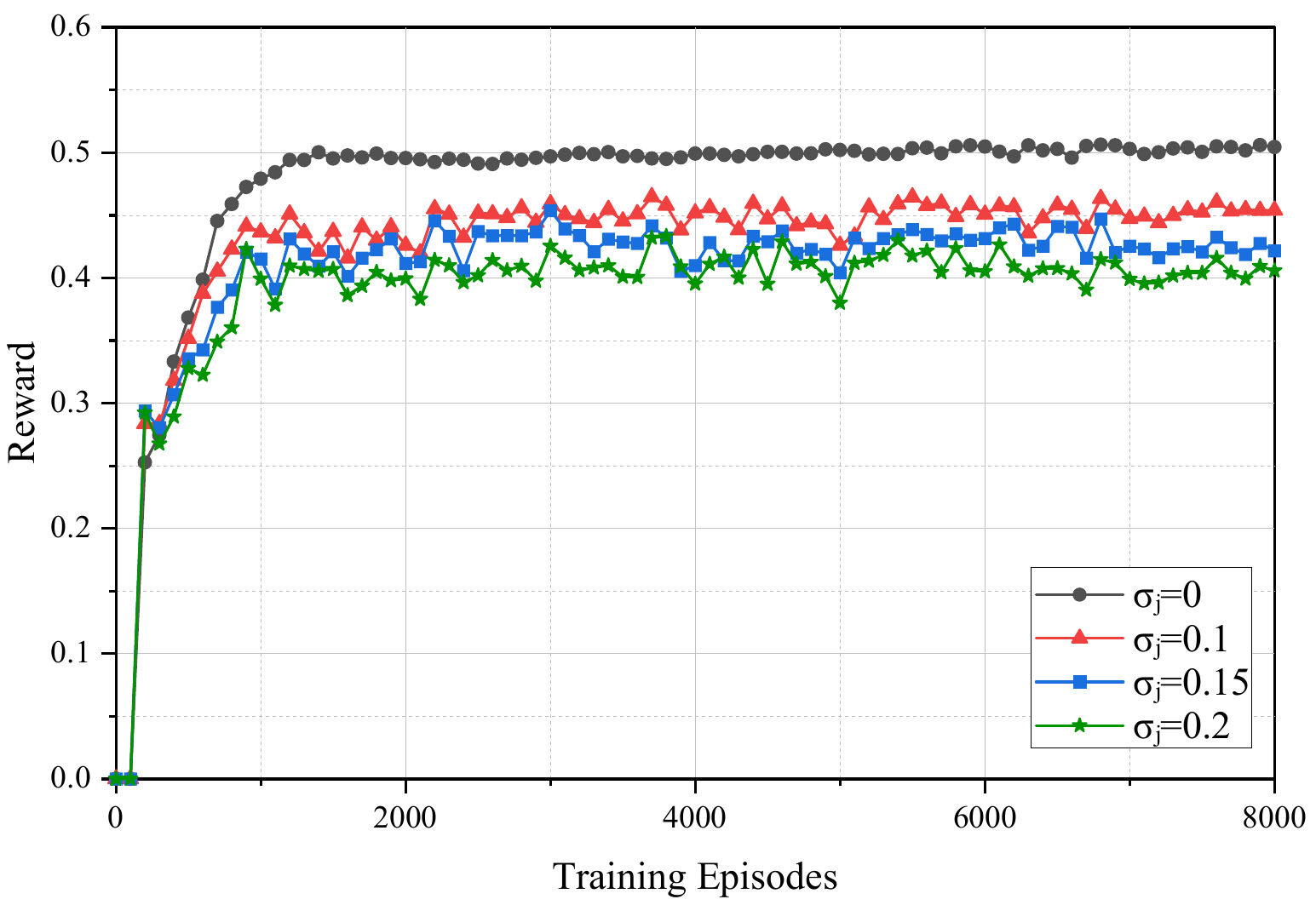}
\caption{Cumulative reward per training episode under varying UAV angular jitter levels.}
\label{Fig_Rewards}
\end{figure}
\begin{figure}[!t]
\centering
\includegraphics[width=0.7\columnwidth]{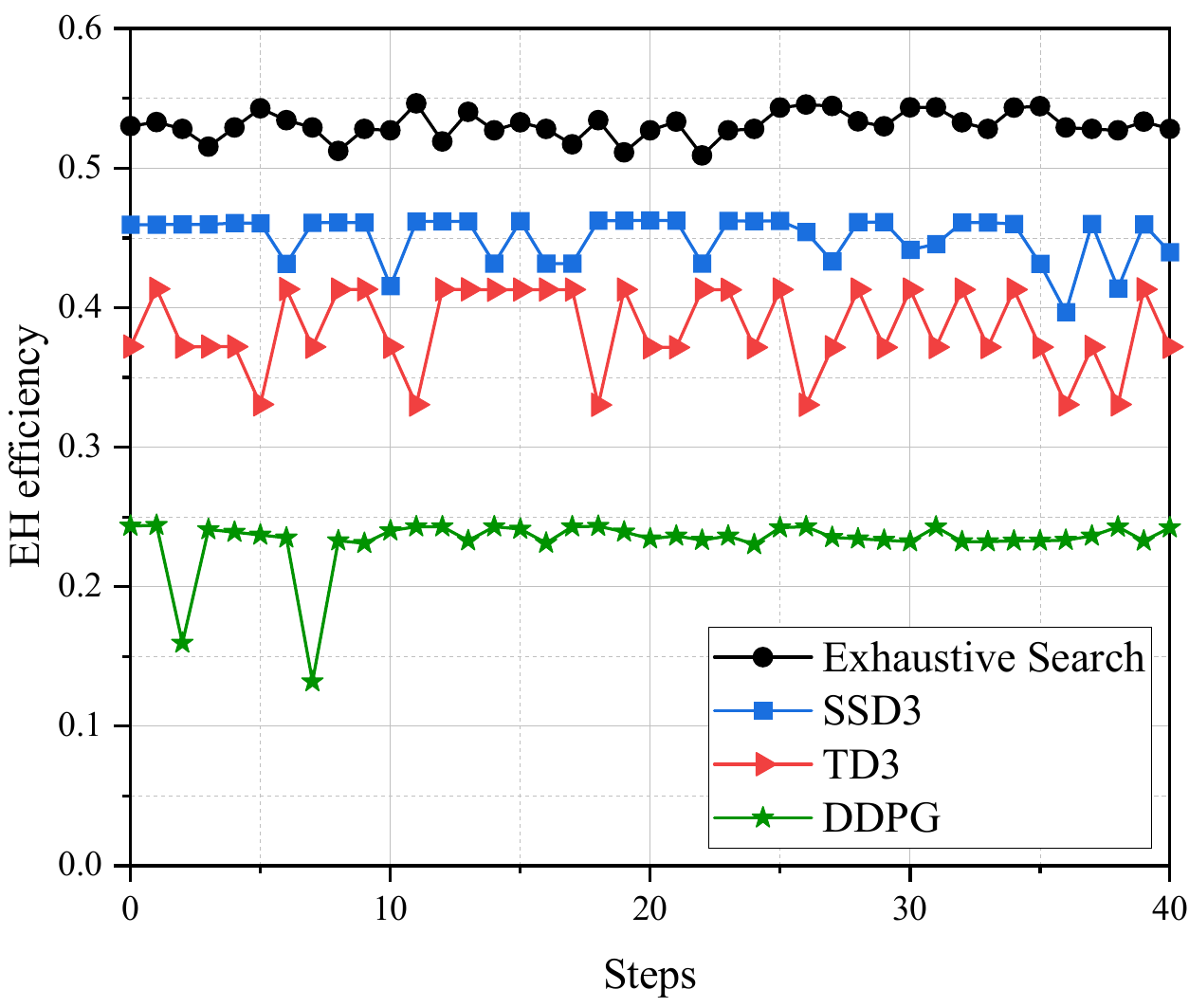}
\caption{Comparison of EH efficiency per time step across different learning-based and baseline algorithms with $\sigma_j=0.1$.}
\label{Fig_EH}
\end{figure}
Fig. \ref{Fig_Rewards} illustrates the convergence of the proposed SSD3 algorithm under varying levels of UAV angular jitter. As expected, performance degrades with increasing jitter variance, but the cumulative reward consistently converges across all cases, demonstrating the robustness of SSD3 to orientation instability. Notably, the gap between the no-jitter case and higher-jitter scenarios (e.g., $\sigma_j=$ 0.2) remains bounded, confirming the algorithm's resilience in realistic UAV conditions.

Fig. 3 compares the EH efficiency of SSD3 with DDPG, TD3, and exhaustive search under a jitter level of $\sigma_j=$ 0.1. While SSD3 achieves slightly lower EH efficiency than the exhaustive search, it provides a significantly lower-complexity and more intelligent and adaptive solution suited for real-time UAV-RIS system operation. Compared to DDPG and TD3, SSD3 consistently delivers higher performance across all time steps, benefiting from entropy regularization and softmax-weighted Q-value estimation that enhance stability and exploration.

\begin{table}[!t]
\centering
\caption{Average EH Efficiency for $\sigma_j = 0.1$}
\scalebox{0.8}{
\begin{tabular}{|c|c|}
\hline
\textbf{Algorithm} & \textbf{Average EH Efficiency (\%)} \\
\hline
Exhaustive Search & 53.09 \\ \hline
SSD3 (Proposed)   & 45.07 \\ \hline
TD3               & 38.48 \\ \hline
DDPG              & 23.28 \\ 
\hline
\end{tabular}}
\label{tab:avg_eh_efficiency}
\end{table}

Table~\ref{tab:avg_eh_efficiency} summarizes the average EH efficiency across different optimization schemes for \( \sigma_j = 0.1 \). The proposed SSD3 algorithm achieves an average EH efficiency of 45.07\%, closely approaching that of the exhaustive search method, which yields the highest efficiency at 53.09\% but incurs an extensive computational cost. Compared to SSD3, the TD3 baseline achieves 38.48\%, while DDPG records the lowest performance at 23.28\% due to its known overestimation bias and reduced stability. 

%

\section{Conclusion}
In this letter, we proposed a DRL-based joint optimization framework to maximize the EH efficiency in a UAV-mounted RIS system under nonlinear RF EH dynamics and UAV jitter. The original nonconvex and time-coupled optimization problem involving BS power allocation, RIS phase shifts, and TS control was reformulated as an MDP. To solve it, we developed the SSD3 algorithm, which integrates dual actor-critic networks, entropy regularization, and softmax-weighted Q-value estimation for improved learning stability and exploration. Simulation results demonstrated that SSD3 achieves superior EH efficiency and faster convergence compared to DDPG and TD3 baselines, and remains robust under varying levels of UAV angular jitter. These results confirm that SSD3 offers a practical and effective solution for energy-constrained UAV-mounted RIS systems, achieving near-optimal performance with lower complexity than exhaustive search and greater robustness than conventional DRL methods.

\bibliographystyle{IEEEtran}
\bibliography{Main}

\newpage

 




\vfill

\end{document}